\documentclass[a4paper,11pt]{article}
\usepackage{pos}

\title{Effects of transverse momentum broadening of parton cascades from coherent emissions and scatterings in a medium}
\ShortTitle{$k_T$ broadening effects in jets from coherent emissions and scatterings}

\author[a]{E. Blanco}
\author[a]{K. Kutak}
\author*[a]{M. Rohrmoser}
\author[b]{W. Płaczek}
\author[c]{R. Straka}

\affiliation[a]{Institute of Nuclear Physics, Polish Academy of Sciences,\\
ul. Radzikowskiego 152, 31-342 Krak\'{o}w, Poland}

\affiliation[b]{Institute of Applied Computer Science, Jagiellonian University,\\
ul.\ \L{}ojasiewicza 11, 30-348 Krak\'ow, Poland}
\affiliation[c]{AGH University of Science and Technology,\\ Krak\'ow, Poland}

\emailAdd{martin.rohrmoser@ifj.edu.pl}
\emailAdd{eti.bla@orange.fr}
\emailAdd{krzysztof.kutak@ifj.edu.pl}
\emailAdd{Wieslaw.Placzek@uj.edu.pl}
\emailAdd{strakrob@gmail.com}

\abstract{Highly energetic particles in the medium of a quark gluon plasma undergo processes of coherent medium induced radiation and scatterings off medium particles. 
We solved the evolution equations for particle fragmentation functions that describe in-medium evolution via these two in-medium processes numerically via Monte-Carlo algorithms. We study the in-medium broadening of the distribution over momentum components transverse to the jet-axis for different cases of jet-medium interactions and different values of average transverse momentum transfer.
}

\FullConference{%
  *** Particles and Nuclei International Conference - PANIC2021 ***\\
  *** 5 - 10 September, 2021 ***\\
  *** Online ***
}

\begin{document}
\maketitle
In this work, we particularly focus on the evolution of parton jets, highly energetic collimated sprays of strongly interacting particles, which can be used as a means to study the hot and dense medium of a quark gluon plasma (QGP).
Within the medium jet particles can undergo processes of scatterings off medium particles.
These scattering processes can also involve particle momenta off the mass shell, which lead to emissions of new particles. 
The emissions can occur simultaneously to multiple scatterings off medium particles giving rise to interference effects. 
The spectra for these coherent medium induced radiations were first described for a QCD medium by Baier, Dokshitzer, Mueller, Peign\'e, Schiff and independently by Zakharov (BDMPS-Z)
~\cite{Baier:2000mf,Baier:2000sb,Zakharov:1996fv,Zakharov:1997uu,Zakharov:1999zk,Baier:1994bd,Baier:1996vi}. 
Later on an effective splitting kernel for coherent medium induced radiations that also allows to reproduce the BDMPS-Z emission spectra was obtained in 
~\cite{Blaizot:2012fh}
as
\begin{equation}
{\cal K}(\mathbf{Q},z,p_0^+)=\frac{2}{p_0^+}\frac{P_{gg}(z)}{z(1-z)}\sin\left[\frac{\mathbf{Q}^2}{2\omega_0\hat q_0}\right]\exp\left[-\frac{\mathbf{Q}^2}{2\omega_0\hat q_0}\right] 
\label{eq:Kqz}
\end{equation}
with
\begin{equation}
\mathbf{Q}=\mathbf{k}-z\,\mathbf{q}\,,\,\, \omega_0=z(1-z)p_0^+\,,\,\,\,\hat q_0=\hat q (1-z(1-z))\,,\,\,\,\, P_{gg}(z)=N_c\frac{\left[1-z(1-z)\right]^2}{z(1-z)},    
\end{equation}
where $p_0^+=E$ is the energy of the initial jet particle, $z$ the parton momentum fraction (with regard to the decaying particle), $\mathbf{k}$ the jet-particle momentum components transverse to the jet axis, $\hat{q}$ the average transverse momentum transfer, $\alpha_S$ the QCD coupling constant and $N_C$ the number of colors. 
An evolution equation for the fragmentation function $D(x,\mathbf{k},t)$ of gluons that includes besides coherent medium induced radiations also scatterings has been formulated 
~\cite{Blaizot:2013vha,Blaizot:2014rla}
as
\begin{eqnarray}
\frac{\partial}{\partial t} D(x,\mathbf{k},t) &=&    \alpha_s \int_0^1 dz\, \int\frac{d^2q}{(2\pi)^2}\left[2{\cal K}(\mathbf{Q},z,\frac{x}{z}p_0^+) D\left(\frac{x}{z},\mathbf{q},t\right) 
- {\cal K}(\mathbf{q},z,xp_0^+) D(x,\mathbf{k},t) \right] \nonumber\\
&+& \int \frac{d^2\mathbf{l}}{(2\pi)^2} \,w(\mathbf{l})D(x,\mathbf{k}-\mathbf{l},t) -  \int \frac{d^2\mathbf{l'}}{(2\pi)^2}\,w(\mathbf{l'})\, D(x,\mathbf{k},t).
\label{eq:BDIM1}
\end{eqnarray}
For our current work we used the following scattering kernels from~\cite{Blaizot:2013vha,Blaizot:2014rla} and  \cite{Aurenche:2002pd}
\begin{align}
 w(\mathbf{l}) &= \frac{16\pi^2\alpha_s^2N_cn}{\mathbf{l}^4}\,,\qquad \textrm{and}\label{eq:wq1}
\\
 w(\mathbf{l}) &= \frac{4\pi\alpha_s m_D^2T}{\mathbf{l}^2(\mathbf{l}^2+m_D^2)}\,.
\label{eq:wq2}
\end{align}
where $n$ is the density of scatterers in the medium
and $m_D$ the Debye mass.

\begin{figure}[!ht]
\centering{}
\includegraphics[width=0.32\textwidth]{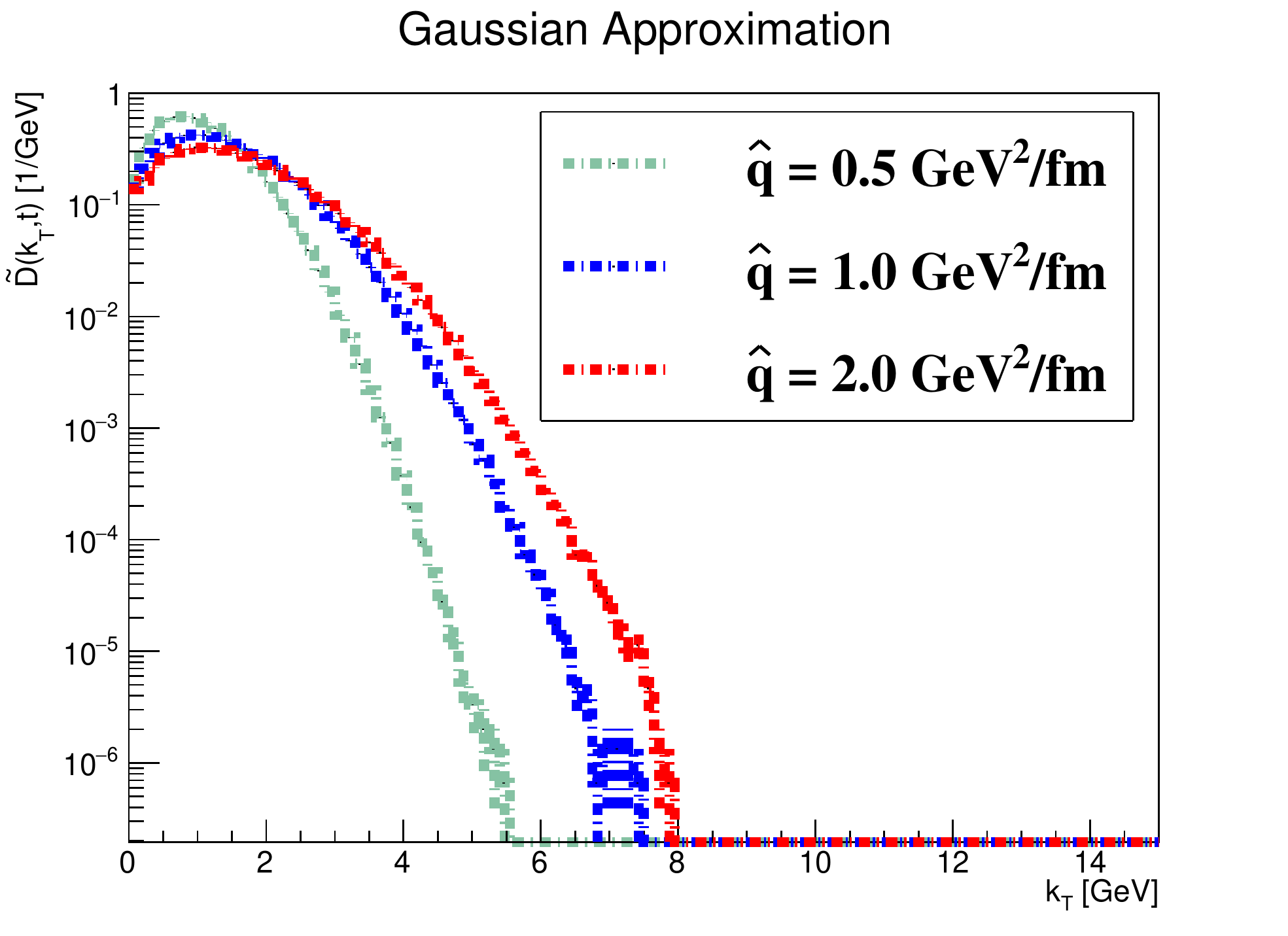}
\includegraphics[width=0.32\textwidth]{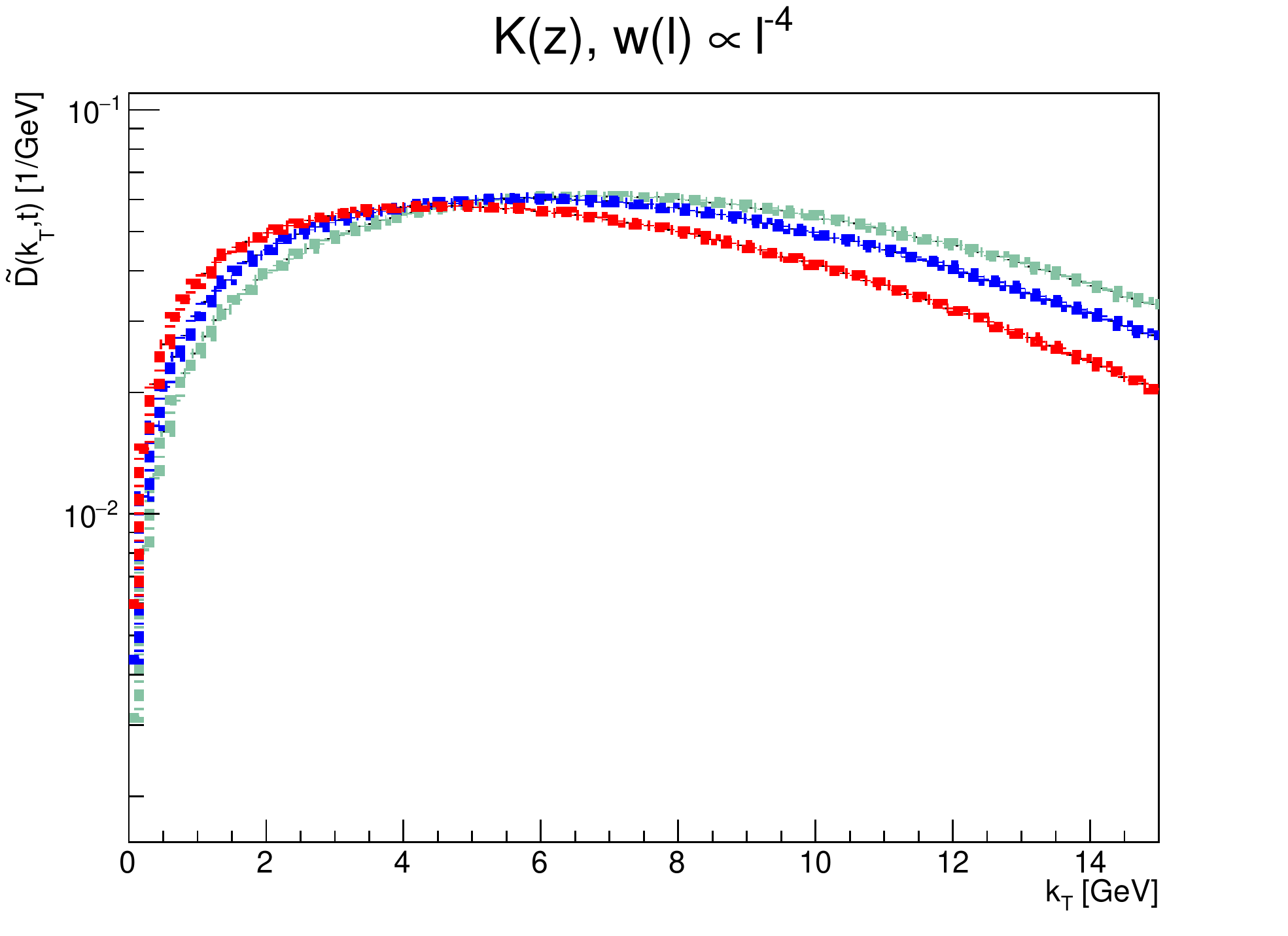}
\includegraphics[width=0.32\textwidth]{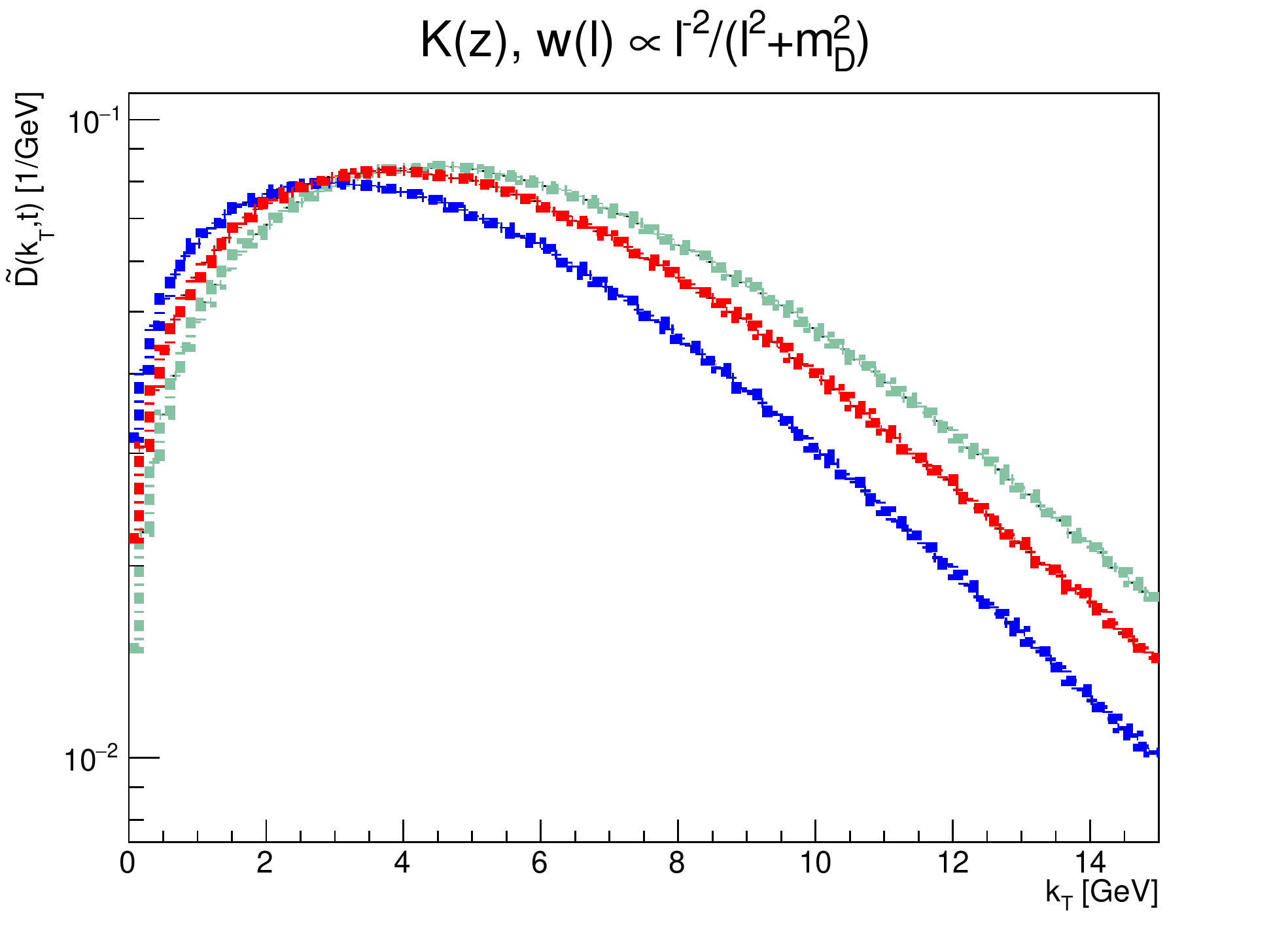}
\includegraphics[width=0.32\textwidth]{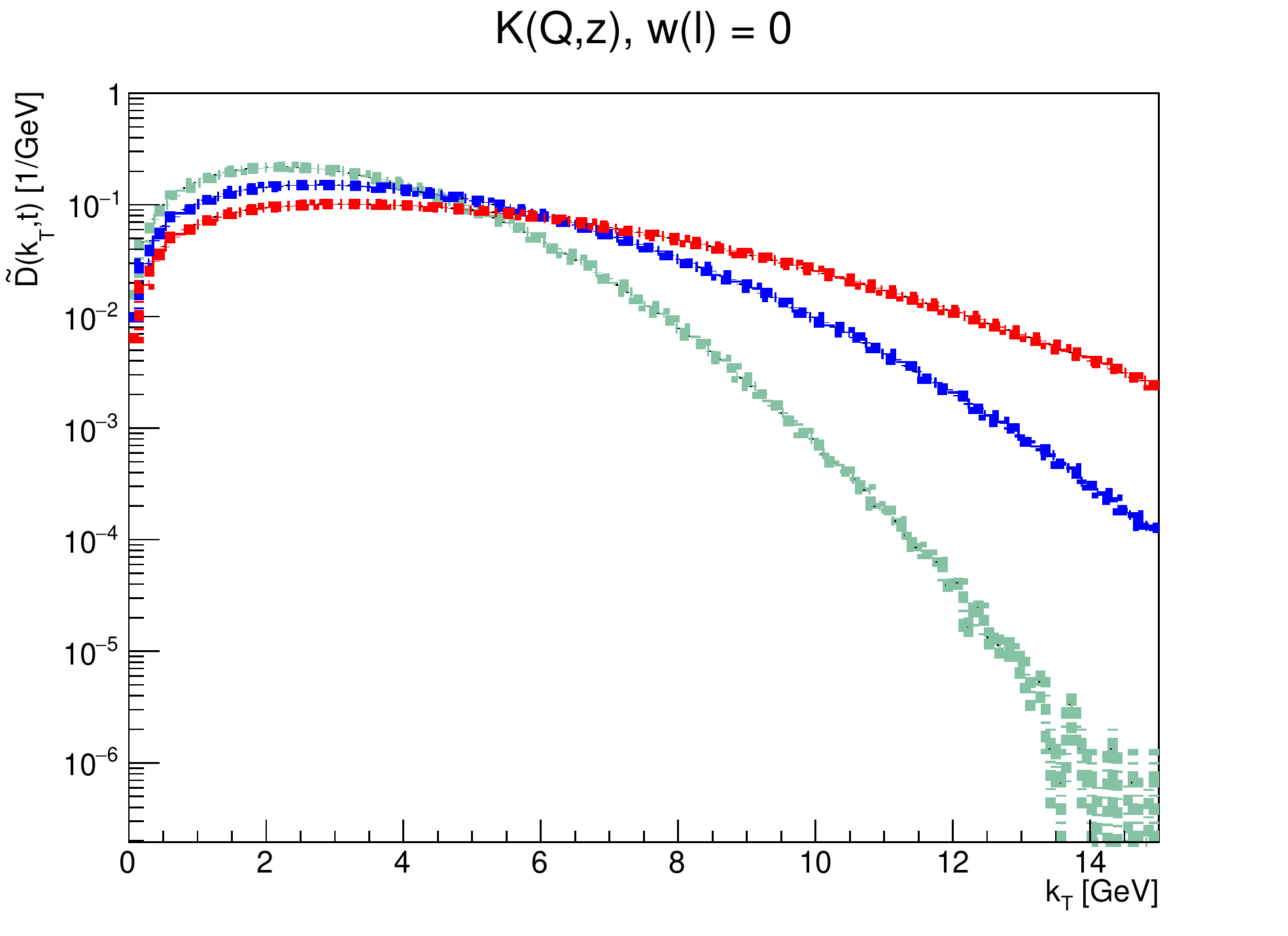}
\includegraphics[width=0.32\textwidth]{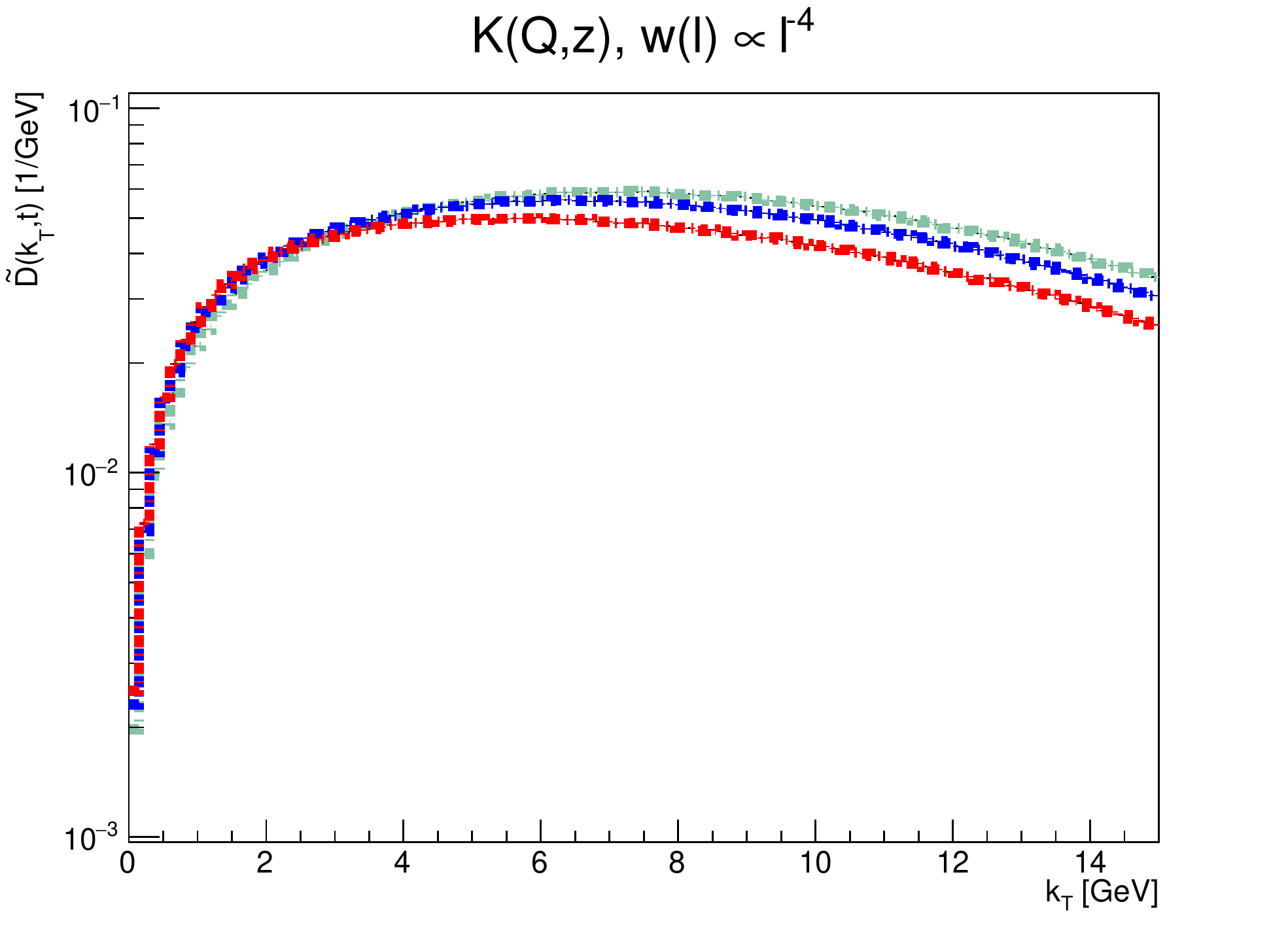}
\includegraphics[width=0.32\textwidth]{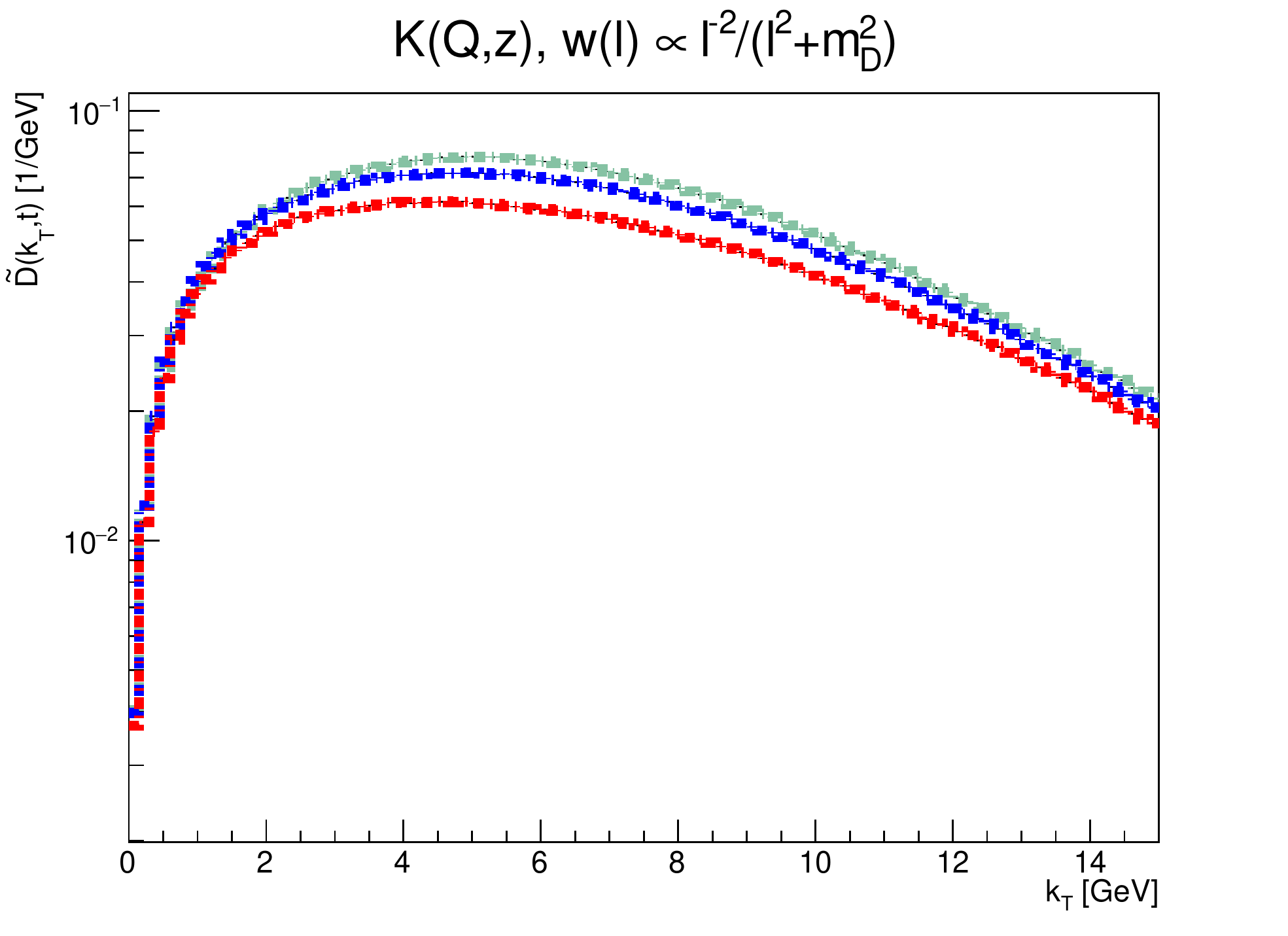}
\caption{Transverse momentum distributions for varying values of $\hat{q}$ in different models as indicated.}
\label{Fig1}
\end{figure}

Via integration over the transverse momentum components an effective collinear splitting kernel $\mathcal{K}(z)$ can be obtained as 
    \begin{equation}
       {\cal K}(z)=\int d^2\mathbf{Q}  {\cal K}(\mathbf{Q},z,yp_0^+)\frac{\sqrt{y}}{2\pi}t^\ast\frac{\alpha_s N_c}{\pi}=\frac{(1-z(1-z))^{5/2}}{(z(1-z))^{3/2}}\,, \qquad \textrm{with }{t^\ast}=\frac{\pi}{\alpha_s N_c}\sqrt{\frac{p_0^+}{\hat{q}}}\,.
       \label{eq:fromKzQtoKz}
    \end{equation}
which corresponds to the following evolution equation~\cite{Blaizot:2013vha,Blaizot:2014rla}
    \begin{eqnarray}
\frac{\partial}{\partial t} D(x,\mathbf{k},t) &=& \frac{1}{t^*} \int_0^1 dz {\cal K}(z) \left[\frac{1}{z^2}\sqrt{\frac{z}{x}} D\left(\frac{x}{z},\frac{\mathbf{k}}{z},t\right)\theta(z-x) 
- \frac{z}{\sqrt{x}} D(x,\mathbf{k},t) \right] \nonumber\\&+&\int \frac{d^2\mathbf{l}}{(2\pi)^2} \,w(\mathbf{l})D(x,\mathbf{k}-\mathbf{l},t)-  \int \frac{d^2\mathbf{l'}}{(2\pi)^2}\,w(\mathbf{l'}) D(x,\mathbf{k},t).
\label{eq:BDIM2}
\end{eqnarray}
After integration over the transverse momentum an evolution equation for the fragmentation functions integrated over transverse momentum $D(x,t)$ can be obtained as
    \begin{equation}
\frac{\partial}{\partial t} D(x,t) =  \: \frac{1}{t^*} \int_0^1 dz\, {\cal K}(z) \left[\sqrt{\frac{z}{x}}\, D\left(\frac{x}{z},t\right)\theta(z-x) 
- \frac{z}{\sqrt{x}}\, D(x,t) \right]\,.
\label{eq:BDIM_coll}
\end{equation}
For the here presented work, we solved the evolution equations~(\ref{eq:BDIM1}),~(\ref{eq:BDIM2}), and~(\ref{eq:BDIM_coll}) numerically via Monte-Carlo algorithms~\cite{Kutak:2018dim,Blanco:2020uzy}.
The following six cases of in-medium jet-evolution were studied: three cases with non-collinear branching, which follow Eq.~(\ref{eq:BDIM1}), where the scattering kernels either follow Eq.~(\ref{eq:wq1}), Eq.~(\ref{eq:wq2}) or $w(q)=0$, two cases with collinear branching that follow Eq.~(\ref{eq:BDIM1}), where the scattering kernels either follow Eq.~(\ref{eq:wq1}), Eq.~(\ref{eq:wq2}), and a case that will be referred to as Gaussian approximation.
In the Gaussian approximation $D(x,\mathbf{k},t)$ is given by 
\begin{equation}
D(x,\mathbf{k},t) = D(x,t)\,\frac{4\pi}{\langle k_\perp^2\rangle}
\exp\left[-\frac{\mathbf{k}^2}{\langle k_\perp^2\rangle}\right],\qquad \textrm{with }\langle k_\perp^2\rangle=\min\left\{\frac{1}{2}\hat q t(1+x^2),\, \frac{x E \hat q}{4\bar\alpha},
\,(x E)^2\right\}\,,
\end{equation}
 where $D(x,t)$ follows Eq.~(\ref{eq:BDIM_coll}) and it is assumed that $k_\perp^2<\omega^2=(xE)^2$.
These models were used to study the relative importance of non-collinear branching and scattering for transverse momentum broadening. 
For the numerical calculations a few constant parameters were assumed as well as a medium described by constant parameters. These parameters are 
\begin{equation}
\hat{q}=1 \textrm{GeV}^2/\textrm{fm}\,,\,\,\alpha_s=\frac{\pi}{10}\,,\,\,
n= 0.243\textrm{GeV}^3\,,\,\,
m_D= 0.993\textrm{GeV}\,,\,\,
p_0^+=100\textrm{GeV}\,,\,\,t=4\textrm{fm/c}\,,
\end{equation}
where $t$ is the time that the jets need to traverse the medium.
We studied the dependence of the distribution 
$
\Tilde{D}(x,k_T,t) = \int_0^{2\pi} 
    d \phi\,k_T\,D(x,\mathbf{k},t)$, with $k_T=||\mathbf{k}||$ on varying values of the average transverse momentum transfer $\hat{q}$.
Results are shown in Fig.~\ref{Fig1}.
It can be seen that for the Gaussian model and the model with non-collinear branching without scattering the distributions in transverse momentum become broader for increasing values of $\hat{q}$.
However, for cases with both branching as well as scattering the distributions for the smallest $\hat{q}$ value, $\hat{q}=0.5$~GeV$^3$/fm, is the broadest, indicating a non-trivial dependence on the interplay between broadenings and scatterings.

\acknowledgments
This work was partially supported by the Polish National Science Centre with the grant no.\ DEC-2017/27/B/ST2/01985.
\bibliographystyle{JHEP}
\bibliography{refs1}

\providecommand{\href}[2]{#2}\begingroup\raggedright\begin{thebibliography}{10}

\bibitem{Baier:2000mf}
R.~Baier, D.~Schiff and B.G.~Zakharov, \emph{{Energy loss in perturbative
  QCD}}, \href{https://doi.org/10.1146/annurev.nucl.50.1.37}{\emph{Ann. Rev.
  Nucl. Part. Sci.} {\bfseries 50} (2000) 37}
  [\href{https://arxiv.org/abs/hep-ph/0002198}{{\ttfamily hep-ph/0002198}}].

\bibitem{Baier:2000sb}
R.~Baier, A.H.~Mueller, D.~Schiff and D.T.~Son, \emph{{'Bottom up'
  thermalization in heavy ion collisions}},
  \href{https://doi.org/10.1016/S0370-2693(01)00191-5}{\emph{Phys. Lett. B}
  {\bfseries 502} (2001) 51}
  [\href{https://arxiv.org/abs/hep-ph/0009237}{{\ttfamily hep-ph/0009237}}].

\bibitem{Zakharov:1996fv}
B.G.~Zakharov, \emph{{Fully quantum treatment of the Landau-Pomeranchuk-Migdal
  effect in QED and QCD}}, \href{https://doi.org/10.1134/1.567126}{\emph{JETP
  Lett.} {\bfseries 63} (1996) 952}
  [\href{https://arxiv.org/abs/hep-ph/9607440}{{\ttfamily hep-ph/9607440}}].

\bibitem{Zakharov:1997uu}
B.G.~Zakharov, \emph{{Radiative energy loss of high-energy quarks in finite
  size nuclear matter and quark - gluon plasma}},
  \href{https://doi.org/10.1134/1.567389}{\emph{JETP Lett.} {\bfseries 65}
  (1997) 615} [\href{https://arxiv.org/abs/hep-ph/9704255}{{\ttfamily
  hep-ph/9704255}}].

\bibitem{Zakharov:1999zk}
B.G.~Zakharov, \emph{{Transverse spectra of radiation processes in-medium}},
  \href{https://doi.org/10.1134/1.568149}{\emph{JETP Lett.} {\bfseries 70}
  (1999) 176} [\href{https://arxiv.org/abs/hep-ph/9906536}{{\ttfamily
  hep-ph/9906536}}].

\bibitem{Baier:1994bd}
R.~Baier, Y.L.~Dokshitzer, S.~Peigne and D.~Schiff, \emph{{Induced gluon
  radiation in a QCD medium}},
  \href{https://doi.org/10.1016/0370-2693(94)01617-L}{\emph{Phys. Lett. B}
  {\bfseries 345} (1995) 277}
  [\href{https://arxiv.org/abs/hep-ph/9411409}{{\ttfamily hep-ph/9411409}}].

\bibitem{Baier:1996vi}
R.~Baier, Y.L.~Dokshitzer, A.H.~Mueller, S.~Peigne and D.~Schiff, \emph{{The
  Landau-Pomeranchuk-Migdal effect in QED}},
  \href{https://doi.org/10.1016/0550-3213(96)00426-9}{\emph{Nucl. Phys. B}
  {\bfseries 478} (1996) 577}
  [\href{https://arxiv.org/abs/hep-ph/9604327}{{\ttfamily hep-ph/9604327}}].

\bibitem{Blaizot:2012fh}
J.-P.~Blaizot, F.~Dominguez, E.~Iancu and Y.~Mehtar-Tani, \emph{{Medium-induced
  gluon branching}}, \href{https://doi.org/10.1007/JHEP01(2013)143}{\emph{JHEP}
  {\bfseries 01} (2013) 143} [\href{https://arxiv.org/abs/1209.4585}{{\ttfamily
  1209.4585}}].

\bibitem{Blaizot:2013vha}
J.-P.~Blaizot, F.~Dominguez, E.~Iancu and Y.~Mehtar-Tani, \emph{{Probabilistic
  picture for medium-induced jet evolution}},
  \href{https://doi.org/10.1007/JHEP06(2014)075}{\emph{JHEP} {\bfseries 06}
  (2014) 075} [\href{https://arxiv.org/abs/1311.5823}{{\ttfamily 1311.5823}}].

\bibitem{Blaizot:2014rla}
J.-P.~Blaizot, L.~Fister and Y.~Mehtar-Tani, \emph{{Angular distribution of
  medium-induced QCD cascades}},
  \href{https://doi.org/10.1016/j.nuclphysa.2015.03.014}{\emph{Nucl. Phys.}
  {\bfseries A940} (2015) 67}
  [\href{https://arxiv.org/abs/1409.6202}{{\ttfamily 1409.6202}}].

\bibitem{Aurenche:2002pd}
P.~Aurenche, F.~Gelis and H.~Zaraket, \emph{{A Simple sum rule for the thermal
  gluon spectral function and applications}},
  \href{https://doi.org/10.1088/1126-6708/2002/05/043}{\emph{JHEP} {\bfseries
  05} (2002) 043} [\href{https://arxiv.org/abs/hep-ph/0204146}{{\ttfamily
  hep-ph/0204146}}].

\bibitem{Kutak:2018dim}
K.~Kutak, W.~P\l{}aczek and R.~Straka, \emph{{Solutions of evolution equations
  for medium-induced QCD cascades}},
  \href{https://doi.org/10.1140/epjc/s10052-019-6838-9}{\emph{Eur. Phys. J. C}
  {\bfseries 79} (2019) 317}
  [\href{https://arxiv.org/abs/1811.06390}{{\ttfamily 1811.06390}}].

\bibitem{Blanco:2020uzy}
E.~Blanco, K.~Kutak, W.~P\l{}aczek, M.~Rohrmoser and R.~Straka, \emph{{Medium
  induced QCD cascades: broadening and rescattering during branching}},
  \href{https://doi.org/10.1007/JHEP04(2021)014}{\emph{JHEP} {\bfseries 04}
  (2021) 014} [\href{https://arxiv.org/abs/2009.03876}{{\ttfamily
  2009.03876}}].

\end{thebibliography}\endgroup

\end{document}